# Evolution of nematic and ferromagnetic ordering in suspensions of magnetic nanoplatelets

Alenka Mertelj,*[a] Borut Lampret,[a,b] Darja Lisjak,[a] Jürgen Klepp,[c] Joachim Kohlbrecher,[d] and Martin Čopič [a]

Suspensions of magnetic nanoplatelets in isotropic solvents are very interesting examples of ferrofluids. It has been shown that above a certain concentration $\Phi_{NI}$ such suspensions form a ferromagnetic nematic phase, which makes this system a unique example of a dipolar fluid. The formation of a nematic phase is driven by anisotropic electrostatic and long-range dipolar magnetic interactions. Here, we present studies of the evolution of short range positional and orientational magnetic order in the suspensions with volume fractions below and above $\Phi_{NI}$, using small angle neutron scattering (SANS). The results show that in the absence of an external magnetic field, short range positional and orientational order already exist at relatively low volume fractions. Polarized SANS revealed that the contribution of ferromagnetic ordering to the formation of the nematic phase is significant. The ferromagnetic correlations can be qualitatively explained by a simple model, which takes into account anisotropic screened electrostatic and dipolar magnetic interactions.

## Introduction

Ferrofluids are suspensions of ferro/ferrimagnetic single domain nanoparticles in isotropic solvents and are well known for the magnetoviscous effect and fascinating surface instabilities[1]. Macroscopically they can be described as superparamagnetic liquids, in which external magnetic fields induce magnetization. Magnetic particles in usual ferrofluids have spherical shape. Recently, it has been shown that in a special kind of ferrofluid, in which magnetic constituents are single domain magnetic platelets, above a certain volume fraction a ferromagnetic nematic phase forms[2]. In this phase, the platelets orient on average in the same direction, as it is usual for the nematic phase and, in addition, also platelets' magnetic moments on average orient in the same direction, which results in stable spontaneous magnetization of the suspension[2] even in the absence of an external magnetic field.

The existence of polar, i.e. ferromagnetic or ferroelectric ordering, in dipolar liquids has been a longstanding open question. Some of the models predicted it some not[3,4]. The first question was whether dipolar interaction could lead to long-range ferromagnetic ordering at all. In solid ferromagnetic materials, ferromagnetic phase appears as a result of exchange interaction between the spins. This interaction is in its nature isotropic, the anisotropy comes as correction due to the coupling of the electrons in d-orbitals and, to the crystal structure[5]. Magnetic dipolar interaction is, however, highly anisotropic, and it has been shown experimentally that in a solid magnetic nanoparticles array[6] it can also lead to a ferromagnetic phase. It is interesting that the magnetization and local crystalline structure in such an array were not correlated[6]. The second question was, whether dipolar interaction can lead to a long-range ferromagnetic order in a liquid. The difference between liquid and solid state is, that in the liquid the particles move and they are preferentially located at the position of the local minimum of the energy, which is reflected in positional correlations between the particles. In the case of dipolar interaction, the preferred orientation of the particle will depend on the position of the minimum, so the positional and orientational degrees of freedom cannot be decoupled. And, as it was pointed out by Morozov[7], short-range orientationaly dependent correlations are responsible for the transition to polar phase and should be considered carefully. For example, the effect of positional correlations on the polar order has been studied theoretically in a positionally frozen dipolar system, and it has been shown that positional correlations strongly influence the appearance of polar phases[8].

The orientational correlations between neighbouring particles occur also because of their shape, which at higher concentrations leads to the nematic phase. At higher concentrations, the platelets can form so-called discotic nematic phase, which has been observed in several colloidal suspensions, e.g. suspensions of clays[9–13] and graphene oxide flakes[14].

In suspensions of magnetic platelets, it seems to be clear that both the magnetic interaction and the shape promote orientational ordering, so the question is, which one plays the leading role and what is the interplay between the two. To gain

[a.] J. Stefan Institute, SI-1000 Ljubljana, Slovenia.
[b.] Faculty of Mathematics and Physics, University of Ljubljana, Slovenia
[c.] Faculty of Physics, University of Vienna, A-1090 Vienna, Austria.
[d.] Laboratory for Neutron Scattering and Imaging, PSI, CH-5232 Villigen PSI, Switzerland.
† Footnotes relating to the title and/or authors should appear here.
Electronic Supplementary Information (ESI) available: [details of any supplementary information available should be included here]. See DOI: 10.1039/x0xx00000x

some insight we studied positional and orientational correlations in suspensions of magnetic platelets with different concentrations by small angle neutron scattering (SANS). Using polarized neutrons, information on magnetic correlations in the system was also obtained. In the first part of the paper, the experimental setup is described and the SANS results are presented. In the discussion, the results of numerical calculations of the average interaction between two platelets are shown and compared with the experimental results. At the end, conclusions are given and open questions are briefly addressed.

## Experimental

### Observation of suspensions by polarizing microscopy (POM)

The suspensions of Barium hexaferrite (BaHF) nanoplatelets (Fig. 1a) in t-butanol (see Materials and Methods) with volume concentrations $\Phi$ of 0.02, 0.04, 0.07 and 0.1 were filled between two quartz glass plates, so that the thickness of the suspension layer was either 100 μm or 50 μm (for the suspension with $\Phi$ = 0.1).

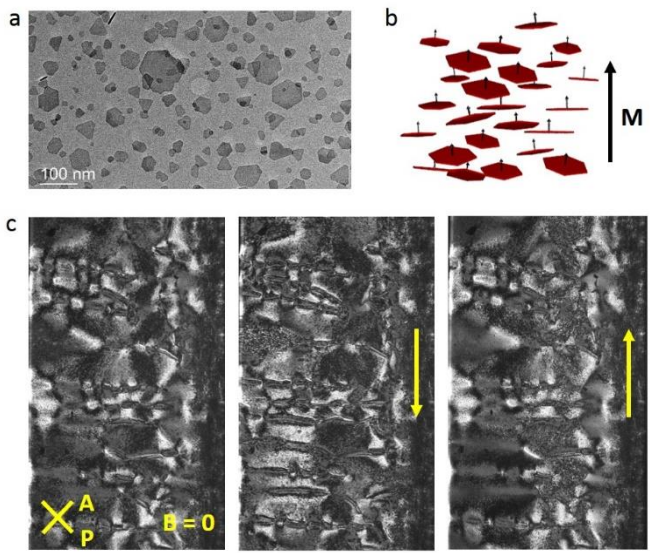

Figure 1: Ferromagnetic nematic suspension. (a) TEM image of BaHF nanoplatelets. (b) Schematic of ferromagnetic nematic phase with average orientation of platelets' magnetic moments along **M**. (c) POM images of magnetic domains in B = 0 (left), in B ~ 20 μT applied in vertical direction down (middle) and up (right) as denoted by the yellow arrows. The orientations of the polarizer (P) and the analyzer (A) was as shown in the left image. The sample with $\Phi$ = 0.1 was in a rectangular capillary with a width of 1 mm and thickness of 50 μm.

The suspensions were first studied optically using POM. Due to the birefringence of the nematic phase, the suspensions in the nematic phase appear bright between crossed polarizers (Fig. 1c), while isotropic suspensions appear dark. The suspensions with concentrations of 0.02, 0.04 and 0.07, appeared bright between crossed polarizers only when an external magnetic field, which induced birefringence, was applied. We have to note here, that a very small field is sufficient to induce birefringence in more concentrated suspensions. For example,

Earth's magnetic field was enough to induce birefringence in the suspension with volume fraction of 0.07. The zero fields condition was achieved by three pairs of coils. In the suspension with concentration of 0.1, homogenously bright regions were observed also in the absence of an external field. The direction of the magnetization varied with the position either in continuous manner or abruptly at the domain walls. Ferromagnetic ordering was further probed by applying a small magnetic field (of order of 20 μT), which caused some of the regions to become brighter and some darker (Fig. 1c.). Homogeneous response of the regions to external field showed that the regions have spontaneous magnetization and the suspension is in a ferromagnetic nematic state (Fig. 1b). The domain configuration depends on the history of the sample's exposure to the external magnetic field and evolves on the timescale of days.

### Small angle neutron scattering (SANS)

The SANS measurements were carried out at the Swiss spallation neutron source (SINQ) beamline SANS I of the Paul Scherrer Institut in Villigen, Switzerland. The samples were placed vertically in the *xz* plane in a neutron beam incoming along the *y* direction (Fig.2). Magnetic field was applied horizontally (in *z* direction) in the plane of the sample using an electromagnet. A 2D detector was used to the measure scattering intensity. The distance between the sample and the detector was chosen so that the scattering in the desired range of the scattering vectors was measured.

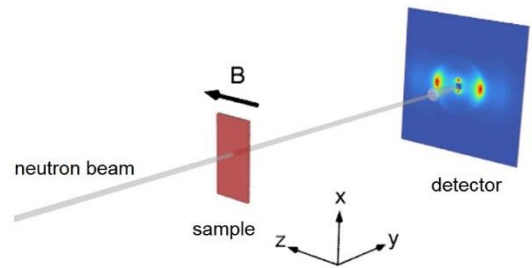

Figure 2: Schematic of a neutron experiment. The sample is placed in a neutron beam incoming along the *y* axis. A 2D detector is placed at some distance from the sample to measure the scattered intensity in the *xz* plane. An external magnetic field is applied horizontally in the *z* direction.

The SANS differential scattering cross section of polydispersed interacting colloids can be written as[15]

$$\frac{d\Sigma}{d\Omega}(\mathbf{q}) = \frac{N_p}{V}\left\langle \left|F(\mathbf{q},\mathbf{n})\right|^2 \right\rangle \\ + V^{-1}\left\langle \sum_{i=1}^{N_p}\sum_{i\neq j} F_i(\mathbf{q},\mathbf{n}_i)F_j(\mathbf{q},\mathbf{n}_j)\exp\left(-i\mathbf{q}\cdot(\mathbf{r}_j-\mathbf{r}_i)\right)\right\rangle \qquad (1)$$

Here, *V* is the scattering volume, $N_p$ the number of the particles within *V*, **q** the scattering vector, $\mathbf{r}_i$ the position and $F(\mathbf{q},\mathbf{n}_i)$ the scattering amplitude of the *i*-th particle. For particles of anisotropic shape, $F(\mathbf{q},\mathbf{n}_i)$ depends also on the particle's orientation denoted by a unit vector $\mathbf{n}_i$. The brackets <> denote averaging over all possible configurations of the particles' positions and orientations at a given temperature, which for an ergodic system is the same as averaging over time.

If the particles are independent, the second term in Eq. (1) is zero and the scattering cross section is a sum of scattering cross sections of individual particles. In this case, the scattering experiment gives information on particles' shape averaged over size and orientation. If correlations between particles exist, then the second term in Eq. (1) gives information on the correlations. In the case of magnetic platelets, the platelets' orientation and relative position are correlated, so the second term in Eq. (1) cannot be decomposed into a product of the structure function and the square average of the form factor as it is usually done for spherical particles[15].

The neutrons are scattered by nuclei and, additionally, their spins interact with the magnetic field, i.e. magnetization of the material, which causes the so-called magnetic scattering[16,17]. So the scattering amplitude consists of two terms, nuclear and magnetic, $F(\mathbf{q},\mathbf{n}) = F_N(\mathbf{q},\mathbf{n}) + F_M(\mathbf{q},\mathbf{n})$. In the case of a particle in a solvent, the nuclear part depends on the contrast between the solvent and the particle $\Delta b$, the particle's volume $V_p$, and its geometrical form factor $f(\mathbf{q})$,

$$F_N(\mathbf{q},\mathbf{n}) = \Delta b V_p f(\mathbf{q},\mathbf{n}) \qquad (2)$$

The scattering amplitude of the magnetic part additionally depends on the direction of the magnetization **M**, and the neutron spin orientation **σ** and polarization, which is reflected in the sign of the magnetic amplitude[18]. For platelets with homogeneous magnetization $M_0$ along the platelet axis **n** in a non-magnetic solvent, it can be written as[18]

$$F_M^{\uparrow\downarrow}(\mathbf{q},\mathbf{n}) = \mp \mu_0 D_M V_p M_0 f(\mathbf{q},\mathbf{n}) \boldsymbol{\sigma} \cdot \left( \frac{\mathbf{q} \times (\mathbf{n} \times \mathbf{q})}{q^2} \right) \qquad (3)$$

Here $D_M = 4.65 \cdot 10^{14}$ (Vs)$^{-1}$. For opposite incoming neutron polarizations (denoted by $\uparrow, \downarrow$), the magnetic part has opposite signs. So by measuring scattering intensities of neutrons with opposite spin, and taking their difference, information on the direction of the magnetization **n** can be obtained:

$$I^\uparrow - I^\downarrow \propto \left\langle |F^\uparrow(\mathbf{q},\mathbf{n})|^2 \right\rangle - \left\langle |F^\downarrow(\mathbf{q},\mathbf{n})|^2 \right\rangle = \left\langle 2 F_N F_M \right\rangle \qquad (4)$$

**SANS**

Figure 3 shows the scattering vector dependence of the scattered intensities measured in diluted suspensions ($\Phi = 0.02$) in the absence of an external magnetic field (Fig. 3a) and in the magnetic field of 2.5 mT (Fig. 3b). In the absence of the field we did not observe any correlations. The line in Fig. 3a is calculated intensity, assuming particles are cylindrical plates with log-normal size distribution (see Materials and Methods). However, as soon as the magnetic field is applied, correlations appear (Fig. 3b). The lines in Figure 3b are again calculated intensities for independent particles in the field of 2.5 mT. For scattering vectors parallel to the applied field the measurements significantly deviate from the model. For **q** || **B**, if the calculated intensity is subtracted from the measurements, we clearly see a peak at $q$ corresponding to the distance of about 40 nm (Inset of Fig. 3b). So the platelets are preferentially located as schematically shown in the inset of Figure 3b. If the field is switched off, correlations disappear again. So, at low concetrations, the field can be used to switch on and off the correlations. This is a behaviour typical for ferrofluids.

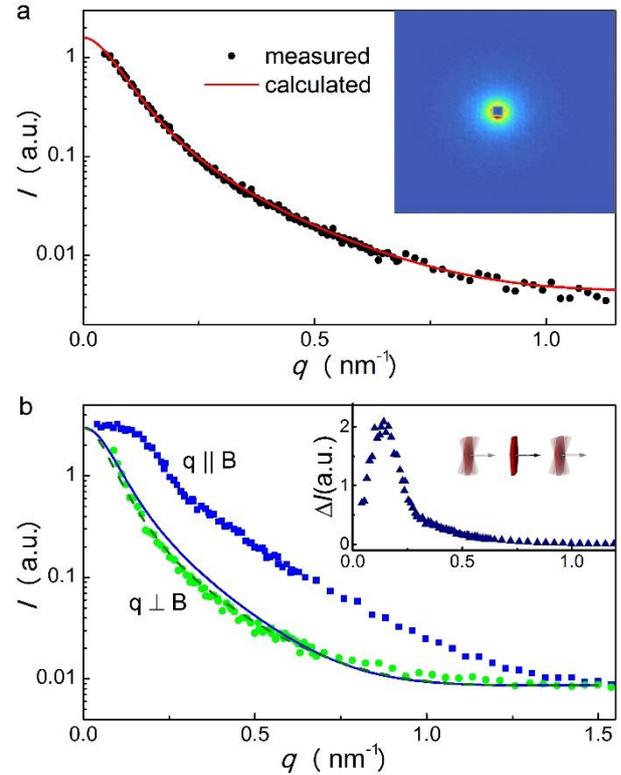

Figure 3: SANS in the suspension with $\Phi = 0.02$. (a) measured (black circles) and calculated (red line) scattered intensity in the absence of an external field. The measured intensity was averaged over azimuthal angles. The inset shows corresponding detector image. (b) scattered intensity along (blue squares) and perpendicularly (green circles) to applied field of 2.5 mT. The lines are calculated scattered intensities (see Materials and Methods) for independent platelets in the field of 2.5 mT for **q** parallel to the field (solid blue line) and perpendicular to the field (dashed olive line). Inset: Difference between measured scattered intensity and calculated intensity for **q** || **B**, which corresponds to the configuration shown schematically in the inset.

In more concentrated suspensions, the local internal field is strong enough for the correlations to be present already in the absence of an external magnetic field (Figure 4). To minimize the contribution of the form factor to the scattering, the measured intensity at higher concentrations $I(\mathbf{q}, \Phi)$ was normalized by the scattering intensity measured at low concentrations $I(\mathbf{q}, \Phi = 0.02)$[15]. Beside a peak at $q_1$ corresponding to a few tens of nm, there is also a secondary peak located at $\sim 2 q_1$. The positions of these peaks depend on the concentrations of the platelets. As expected, $q_1$ increases with the concentration, i.e., higher is the concentration, closer are the platelets. In the nematic phase the distance corresponding to the peaks is about 22 nm. Additionally, peaks get narrower at higher concentrations indicating longer range correlations. Another small peak denoted by $q_2$ is observed at low $q$, corresponding to the distances about 100 nm.

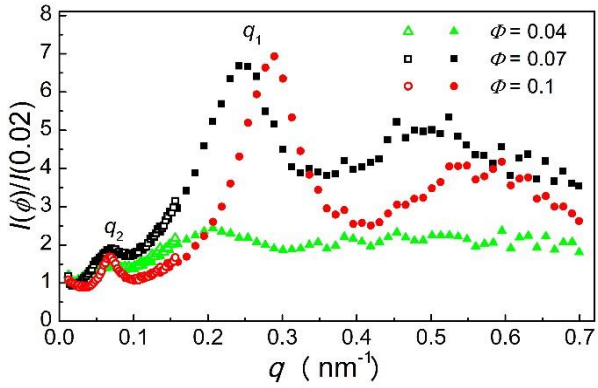

Figure 4: SANS intensity measured in concentrated suspensions in the absence of an external magnetic field normalized by SANS intensity measured in the suspension with $\Phi$ = 0.02 at **B** = 0. The empty symbols present additional measurements that are more precise at small $q$.

In order to gain clearer picture of the positional correlations associated with the observed peaks shown in Fig. 4, a very small field (0.19 mT) was applied to the sample with $\Phi$ = 0.07. The field was small enough not to cause additional correlations, which was checked by comparing the scattering intensity averaged over azimuthal angle with that in the absence of the field. The results clearly show that the peaks denoted with $q_1$ in Figure 4 correspond to correlations in the direction of the field, while the peaks denoted by $q_2$ correspond to correlations perpendicular to it (Fig. 5a). Moreover, using polarized neutrons (SANSPOL) we obtained information on the orientational correlations (Figure 5b). It turns out that the positional and orientational correlations coincide. Furthermore, in the direction of the field the neighbouring platelets are preferentially oriented in the same direction, while when their relative position is in the direction perpendicular to the field they prefer opposite, i.e., antiferromagnetic orientation (Fig. 5c).

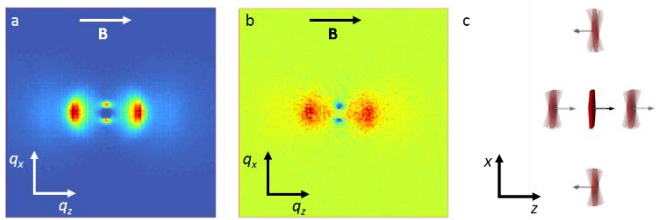

Figure 5: SANSPOL in suspension with $\Phi$ = 0.07. a) Total SANS intensitiy $I^\uparrow + I^\downarrow$ showing positional correlations. The peaks in the horizontal direction correspond to ~30 nm, while the ones in the vertical direction to ~90 nm. b) SANSPOL intensity difference $I^\uparrow - I^\downarrow$ showing magnetic correlations. The blue peaks corresponds to antiferromagnetic correlations, and the red peaks to ferromagnetic correlations. c) Scheme of positional and magnetic correlations of the neighbouring platelets. B = 0.19 mT.

**Field dependence**

In Figure 6a a comparison of the dependence of the scattering intensity on magnetic field between diluted and concentrated suspensions in the isotropic ($\Phi$ = 0.02 and 0.07) and the nematic phase ($\Phi$ = 0.1) is shown. At the beginning of the measurements at B=0 (Fig. 6a, left column), there is little difference between the concentrated suspensions. The scattering intensity in the isotropic suspension is, as expected, azimuthally symmetric. Because the nematic suspension is initially multi-domain, it is also almost azimuthally symmetric. There is only a slight preference in one direction, which may be a consequence of a very small ambient magnetic field. The scattering intensity becomes increasingly azimuthally asymmetric with increasing magnetic field. In the isotropic suspension, the magnetic field induces ordering of the platelets along the field. In the nematic suspension, the platelets within one domain already exhibit homogeneous orientational order, the direction of which is different in different domains. As soon as a small magnetic field is applied, the domains with the magnetization along the field start to grow and their contribution to the scattering becomes dominant. With increasing field average ordering of the platelets also increases.

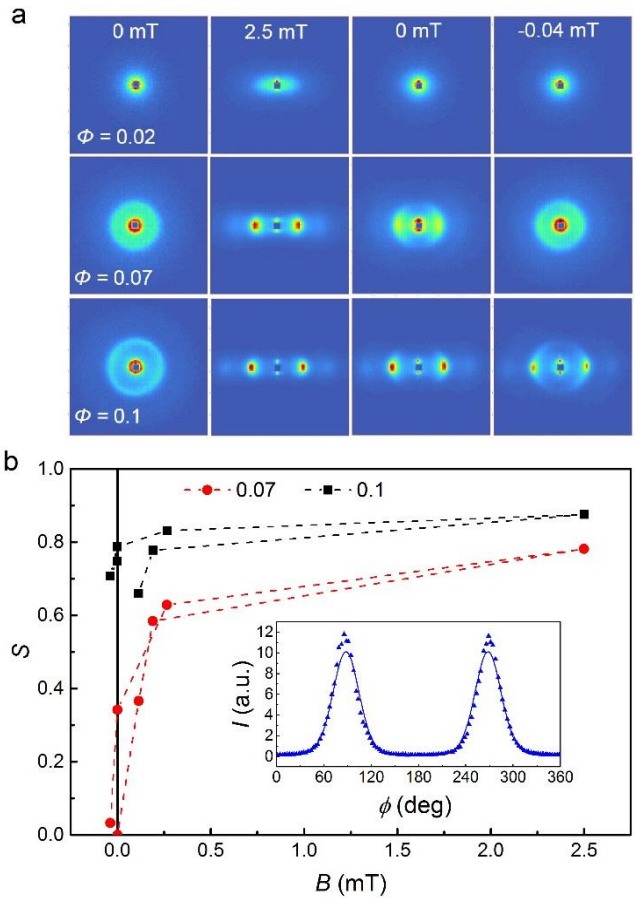

Figure 6: (a) SANS intensity as a function of the external magnetic field applied in the horizontal direction for suspensions with the concentrations as marked. The field was applied in a sequence as marked in the top row. (b) Dependence of the estimated scalar order parameter $S$ (see text) on the external field. Dashed lines are guides to the eye. Inset: an example of a fit used to determine $S$ (see text). Blue triangles are the scattered intensity and the blue line is the fit.

The degree of the ordering in the nematic phase is usually described by the scalar order parameter $S = \langle P_2(\cos\beta) \rangle$ [19], where $P_2(x)$ is the second Legendre polynomial and $\beta$ is the angle between the platelet axis and the average orientation of

                                                                                                                                                                                                                                                          

the nematic liquid crystal. We roughly estimated the degree of ordering as a function of an external field by fitting the dependence of the largest peak ($q_1$) on azimuthal angle (Inset Fig. 6b), using a method for determination of $S$ from SAXS measurements in thermotropic nematic liquid crystals assuming the Maier-Saupe orientational distribution function[20]. Because the platelets are polydispersed and the scattering intensity is weighted by platelets' radius to the 4[th] power, i.e., $R^4$, the order parameter is overestimated, but nevertheless it gives a good insight into the behaviour. In the isotropic suspension, $S$ increases from zero in the absence of the field to around 0.8 in the field 2.5 mT (Fig. 6b). In the nematic suspension, we were not able to determine the order at B=0 as the sample was in a multi-domain state. When the field was applied, as expected, $S$ was larger than that of the isotropic suspension in the same field. The main difference between the isotropic and the nematic suspension was observed, when the field was decreased and switched off (Fig. 6a). In the nematic phase, when we decrease or even switch off the field, the suspension remains ordered. A small field, of the order of Earth's magnetic field, in the opposite direction caused the suspension to become less ordered, however, when we switch it off again, the suspension gets more ordered again. So using an external field we can temporarily magnetize the suspension in the nematic phase. The isotropic suspension also remained slightly oriented, when the field was switched off, however, a small opposite field caused disorientation. The reason for this remaining order is most probably a small change of a remanent field of the electromagnet (of order of a few 0.01 mT).

We mentioned that the width of the peaks in the radial direction gets narrower in more concentrated suspensions, indicating the correlation between more than two platelets. However, the external field does not change the width of the peaks, suggesting that the range of the correlations remains the same.

## Discussion

Due to strong interparticle correlations and long range magnetic interaction, suspensions of magnetic particles present a very challenging system to describe theoretically[3]. In diluted suspensions, the correlations between the particles can be reasonably well described by the interaction between two particles. However, as the concentration of the particles increases many body interactions become more and more important, and, for the description of a long range magnetic ordering, they are crucial.

In order to roughly understand the SANS experiments we calculated the average interaction between two platelets. A magnetic platelet can be approximated by a thin disk with magnetic dipole moment $p_i$ oriented perpendicularly to the plane of the disk, denoted by a unit vector $\mathbf{n}_i$. The main magnetic interaction between two disks is dipolar:

$$U_{dip} = -\frac{\mu_0 p_1 p_2}{4\pi r^3}\left(\frac{3(\mathbf{n}_1 \cdot \mathbf{r})(\mathbf{n}_2 \cdot \mathbf{r})}{\mathbf{r} \cdot \mathbf{r}} - \mathbf{n}_1 \cdot \mathbf{n}_2\right) \quad (5)$$

Here $\mathbf{r}$ is a vector with absolute value $r$ connecting disks' centres and $\mu_0$ is the vacuum permeability. The average magnetic moment of the platelets used in the experiments is ~ $10^{-18}$ Am$^2$, and the average magnitude of the dipolar interaction at $r = 10$ nm has a value $> 25\ k_BT$. This shows that a strong enough repulsive electrostatic interaction between the platelets is crucial for the stabilization of the suspension. The screened electrostatic interaction between two disks with radii $R_i$ and charge $Z_i e_0$ can be approximated by[21]

$$U_{el} = \frac{Z_1 Z_2 e_0^2}{4\pi\varepsilon\varepsilon_0} f(\kappa R_1, \kappa R_2, \mathbf{n}_1, \mathbf{n}_2) \frac{e^{-\kappa r}}{r} \quad (6)$$

Here $\kappa^{-1}$ is the Debye screening length, $\varepsilon$ the solvent's dielectric constant, and $\varepsilon_0$ the vacuum permittivity. Because of the anisotropic shape, the electrostatic interaction is anisotropic, which can be described by the anisotropy function $f$. In the case of very thin disks, $f$ can be approximated by[21]

$$f \approx 4 \frac{I_1(\kappa R_1 \sin\vartheta_1)}{\kappa R_1 \sin\vartheta_1} \frac{I_1(\kappa R_2 \sin\vartheta_2)}{\kappa R_2 \sin\vartheta_2} \quad (7)$$

The angle $\vartheta_i$ is the angle between the i-th disk's orientation $\mathbf{n}_i$ and $\mathbf{r}$, and $I_1(x)$ is the modified Bessel function of the order 1. The charge of the disks can be estimated by assuming that it is similar to the charge of a sphere with equivalent volume (see Materials and Methods). Estimation of $\kappa$ is more difficult. It depends on the concentration of free ions in the suspension, which is difficult to asses correctly. The platelets are positively charged due to the surfactant and, additionally, some of the surfactant is dissolved and partially dissociated in the solvent contributing to free ions, which screen their charge. A portion of surfactant molecules are physisorbed to the platelets' surface[22,23], so that there is a dynamic equilibrium of adsorption and desorption of the surfactant molecules. This equilibrium is affected by the concentration of the surfactant, the concentration of the platelets, and also by the degree of dissociation of the surfactant in the solvent, which all affect the concentration of the free ions and, consequently, $\kappa$. We calculated the average interaction $\langle U_{int} \rangle = \langle U_{dip} + U_{el} \rangle$ between the disks for different screening lengths using the Boltzmann probability distribution (see Materials and Methods) and compared it with the SANS results.

First, we fixed the orientation of the first disk to be along the z axis and calculated the average potential and orientation of the second disk. The size distribution of the platelets' radia is broad, so also the interaction between the platelets varies a lot. For an average disk with $R_{avg}$ = 20 nm and $p_{avg}$ = $10^{-18}$ Am$^2$ (the averages were performed using volume distribution, see Materials and Methods) the interaction $\langle U_{int} \rangle$ is repulsive (Figs. 7a and 7b). This means that strong correlations are not expected in the isotropic phase. However, the volume fraction calculated from the average distances between neighbours as determined from SANS measurements is about a factor of 3 too small, which shows that only larger platelets exhibit correlations, while the small ones contribute to the homogeneous background. So we looked at the interaction between larger disks. We took a portion of the platelets with radia > 23 nm, which constitutes 30% of the volume of all particles. The volume averaged radius and the dipole moment of this portion of the platelets are $R_{30}$ = 31 nm and $p_{30}$ = 2. $10^{-18}$ Am$^2$, respectively. In Fig. 7c the interaction between larger disks is shown for different values of $\kappa$. We see that for $\kappa^{-1}$ below 11 nm the interaction exhibits a minimum in the direction of the dipole of the first disk. The depth and the position of this minimum depend on $\kappa^{-1}$. At $\kappa^{-1}$ = 9 nm (Fig. 7d), it is located at slightly less than 40 nm, which is comparable to the value measured in the experiments.

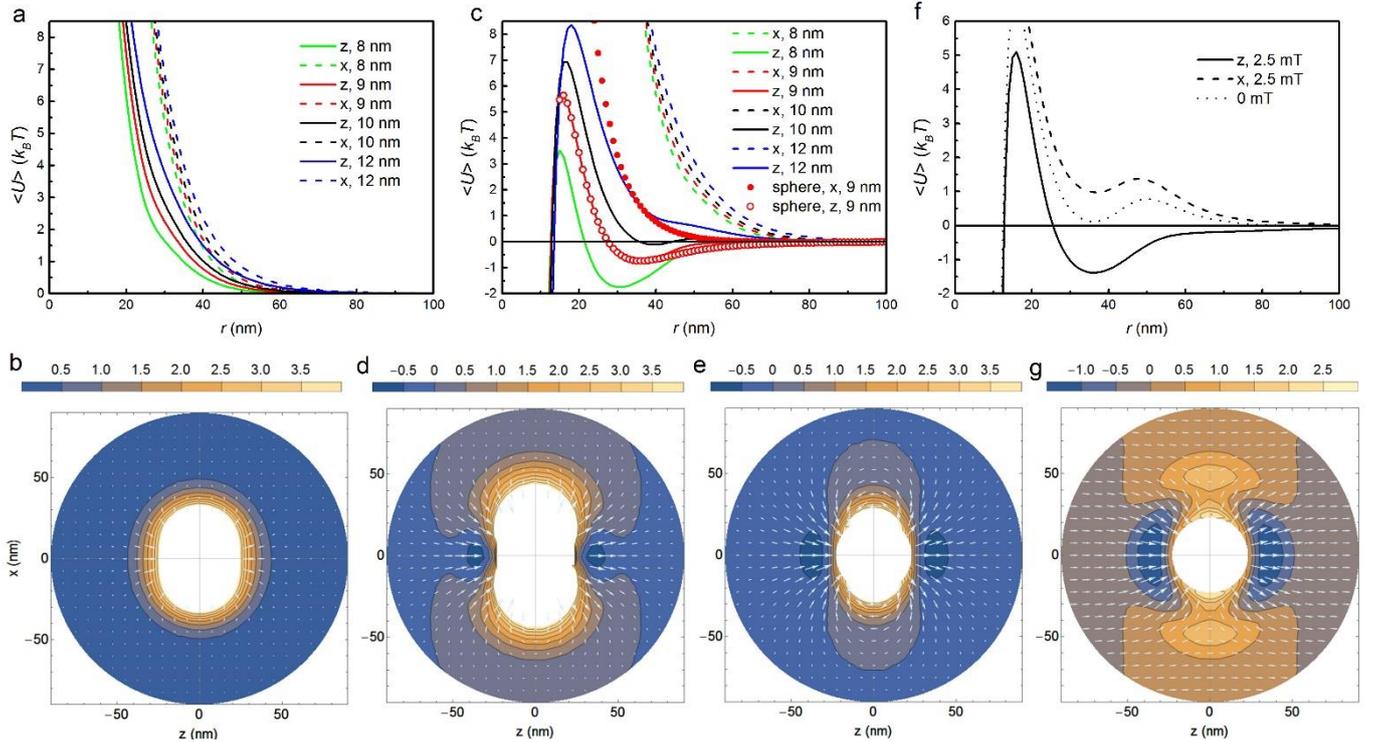

Figure 7: Numerical calculations of the average interaction between two magnetic disks. In a) to e) the orientation of the first disk is fixed along the z axis. a) Interaction between two average disks in the x and z directions for different Debye screening lengths $\kappa^{-1}$ as denoted by the values in the legend. b) Contour plot of the interaction between average disks for $\kappa^{-1}$ = 9 nm. c) Interaction between larger disks (see text) in the x and z direction for different $\kappa^{-1}$ and comparison with a sphere for $\kappa^{-1}$ = 9 nm (full and empty circles). d) Interaction between larger disks for $\kappa^{-1}$ = 9 nm. e) Interaction between two spheres for $\kappa^{-1}$ = 9 nm. f) The average interaction between two larger disks in the absence of the field (dotted line) and in the field of 2.5 mT (solid and dash lines) applied along z axis. g) 2D plot of the average interaction between two larger disks in the field of 2.5 mT for $\kappa^{-1}$ = 9 nm. The arrows in the contour plots denote the average orientation of the second disk and the lengths of the arrows denote the degree of the orientation, which goes from 1 in the minima to almost zero at x or z of 100 nm. The interaction in the contour plots is shown in the units $k_BT$.

To evaluate the influence of the anisotropy of the electrostatic interaction, we also performed the same calculation for a sphere by setting $f$ = 1 (Fig. 7e) and compared it with the interaction between the disks (Fig. 7d). In both cases, there are two minima located in the z direction on each side of the first particle, in which the dipole moment of the second particle is preferentially oriented approximately in the same direction as the first one. However, in the case of the disks, these minima are narrower, which means that there is larger probability that the second particle will be oriented exactly in the same direction if the particles are disks than if they are spheres. In other words, the disk's shape through anisotropic electrostatic interaction promotes ferromagnetic ordering.

Next, we calculated the interaction between the disks in the presence of an external magnetic field **B** (Figure 7f and 7g). In this case, the probability for a given orientation of the disks also depends on the strength of the external field, which is taken into account by additional terms $-\mathbf{p}_1\cdot\mathbf{B}-\mathbf{p}_2\cdot\mathbf{B}$ in the potential used in the Boltzmann probability distribution (Materials and Methods). The averaging over the orientation of both disks is necessary. Because the field induces average orientation of the disks, the minima are deeper than without the field, and the positional and orientational correlations become stronger. This is exactly what we observed in the case of diluted suspension, in which the magnetic field caused correlations (Fig. 3). The depth of the minima depends on the ratio between the electrostatic and magnetic interaction, the ratio between the total interaction and $k_BT$, and the Debye screening length. If the minima are deeper than approximately $k_BT$, flocculation of the platelets would appear, which in diluted suspensions would result in sedimentation, while in concentrated suspensions in formation of a gel.

In comparison with suspensions studied in Ref.[2], where the phase transition to the ferromagnetic nematic phase was observed at a volume fraction of about 0.28 and no positional correlations were observed in the isotropic phase, in the suspensions used in our study, the ferromagnetic nematic phase appeared at much lower volume fractions of 0.1 and strong positional and orientational correlations were observed already in the isotropic phase. As we demonstrated before, a difference in the correlations can be attributed to the difference in the electrostatic interaction between the platelets, for example in the Debye screening length, similarly as the variation of the screened electrostatic interaction affects the phase behaviour of suspensions of gibbsite platelets.[11,24] In magnetic suspensions, strong ferromagnetic correlations lead to a ferromagnetic phase at lower volume fractions of particles. The local field acting on a dipole inside a ferromagnetic fluid can be written as a sum of the Weiss mean field $\mu_0 \mathbf{M}_f/3$, where $\mathbf{M}_f$ is the magnetization of a ferromagnetic fluid, and the contribution of short range correlations.[7] If strong short-range ferromagnetic correlations exist, then a smaller mean field, i.e., smaller volume fraction of magnetic constituents, is needed for appearance of long-range magnetic order.

Our calculations also show that weak antiferromagnetic correlations observed in the experiments cannot be explained by the interaction between two disks. Moreover, comparison of the interactions between pairs of disks and spheres (Figs. 7d and 7e) shows that, in the case of the disk, antiferromagnetic ordering is suppressed. Therefore, for the explanation of the antiferromagnetic correlations many-body interactions have to be taken into account.

The fact that positional and orientational correlations coincide shows that magnetic interaction is very important in formation of the ferromagnetic phase. Our calculations show that the shape can assist its formation and that it is more likely for the disks to form magnetically ordered phase than for the spheres. However, the nematic phase for platelets with the same aspect ratio (thickness/diameter) but without dipole moments is expected to appear at about twice the volume fraction observed in the magnetic suspensions studied here.[25] So the question, whether the shape is crucial for long-range polar order, remains open. Electrostatically stabilized ferrofluids made of spherical magnetic colloids have been investigated by SANS[26,27] and, in concentrated suspensions, strong correlations due to magnetic interaction were also observed in the absence of the field. However, an external magnetic field induced pseudocrystalline ordering, rather than liquid-like ordering.[26]

## Conclusions

The SANS investigation of the evolution of ferromagnetic nematic order in suspensions of magnetic platelets in t-butanol revealed strong positional and orientational correlations in concentrated isotropic and ferromagnetic nematic phases. The orientational correlations are such that they promote ferromagnetic ordering and lead to ferromagnetic nematic phase at much lower concentrations than was previously observed. The results show that only larger particles exhibit correlations, while smaller platelets contribute to homogeneous background. This raises several questions, such as how important is polydispersity for the formation of the ferromagnetic nematic phase; would the phase even exist in suspensions of monodispersed platelets or would it be replaced by columnar nematic phase or even solid columnar phase as was observed in gibbsite suspensions?[9,25] Hexagonal columnar phase, however, would be frustrated from the point of view of magnetic interactions. An open question is also the role of antiferromagnetic correlations. It is expected that in the ferromagnetic nematic phase larger platelets are more ordered than small ones, which is reflected in smaller order parameter $S$ of the small platelets. However, having a portion of platelets orientationaly ordered in the opposite direction is rather unexpected, in particular, because at the distances, at which the antiferromagnetic correlations are observed, the magnetic interaction is much smaller than $k_B T$ even for columns of several platelets. Such unexpected antiparallel ordering of magnetic dipoles has been actually observed in some theoretical approaches, and there the authors raised concerns over the validity of their observation.[7,28] To answer these questions and get a better understanding of the behaviour of the ferromagnetic nematic phase, computer simulations of the system would certainly be very helpful.

## Materials and Methods

### Synthesis and preparation of suspensions

The scandium substituted Ba hexaferite (BaHF) nanoplatelets were synthesized hydrothermally[29] and suspended in t-butanol using surfactant DBSA. The morphology and crystals structure of the platelets was verified with a transmission electron microscope (TEM, Jeol 2100). The thickness of the platelets was predominately 3 nm and 4.1 nm, and 5.3 nm[30,31]. The distribution of the platelet diameter is approximately log-normal, with a mean of 29 nm and a standard deviation of 13 nm as was estimated from equivalent diameter using DigitalMicrograph™ Gatan Inc. software. A minimum of 150 plateletets was accounted in the analysis.

Because of high magnetocrystalline anisotropy of BaHF the magnetic dipole moments of the platelets are perpendicular to the plane of the platelets. The magnetization of the platelets was measured by vibrating sample magnetometer (Lakeshore 7400 Series VMS). The magnetization of the platelets was $1.7 \cdot 10^5$ A/m. The amount of the surfactant was determined thermogravimetrically and was 20.5 wt%.

The volume fraction of the initial suspension was 0.02. The suspension was repeatedly centrifuged at 11200 g to obtain suspensions with higher volume fractions.

### Calculation of scattering intensity for independent disks

The scattering intensity for independent platelets has been calculated by using a form factor of a disk and then averaging $\left\langle \left| F(\mathbf{q},\mathbf{n}) \right|^2 \right\rangle$ over log-normal radius distribution and over orientation distribution, which is constant in the absence of the magnetic field and proportional to $\exp(p\mathbf{n}\cdot\mathbf{B}/k_B T)$, when the field $\mathbf{B}$ is applied. The dipole moment $p$ is proportional to the volume of the disk. In the calculations a constant thickness of 3.5 nm was assumed.

### Numerical calculation of average interaction

The size distribution of the platelets is broad, and since the interaction between two individual platelets depends on their size, the strength of the interaction varies a lot. First, we calculated the interaction between average platelets, which we calculated by averaging using a volume distribution $dV(R)/dR$. It tells the relative volume of platelets with a given $R$, rather than the number distribution, which tells the relative number of the platelets with a given $R$. We chose this averaging because magnetic properties scale with volume.

The charge of a disk, which determines the strength of the electrostatic interaction (Eq. 6), is assumed to be similar to the charge of a sphere with equivalent volume (with radius $R_{eq}$), which is given by[32]

$$Ze_0 = 4\pi\varepsilon\varepsilon_0 \kappa R_{eq}^2 k_B T / e_0 \cdot$$
$$\left(2\sinh(e_0\psi_s/(2k_B T)) + 4/(\kappa R_{eq})\tanh(e_0\psi_s/(4k_B T))\right) \quad (8)$$

In the calculation of $R_{eq}$ the volume of the disks is taken to consist of a magnetic part and a surfactant layer (with the thickness of about 1 nm). The surface potential $\psi_s$ can be taken to have a value of the zeta potential, which was measured to be about 0.075 V.[22]

The average interaction at a given position **r** can be calculated by averaging $U_{int} = U_{dip} + U_{el}$ over orientations of disks assuming the Boltzmann probability distribution $p(\mathbf{n}_1,\mathbf{n}_2,\mathbf{r}) = \frac{1}{A}e^{-U/k_B T}$, where $A$ is the normalization constant $A = \int e^{-U/k_B T} d\Omega_1 d\Omega_2$, and the integration is performed over all solid angles $\Omega_i$. In the case of an external magnetic field, there are additional terms in $U$, $U = U_{int} - \mathbf{p}_1 \cdot \mathbf{B} - \mathbf{p}_2 \cdot \mathbf{B}$, which influences the average orientation. In the absence of an external magnetic field, by fixing the orientation of the first disk to be along the $z$ axis, the average interaction and orientation of the second disk in the field of the first can be calculated.

## Conflicts of interest

There are no conflicts to declare.

## Acknowledgements

The authors acknowledge the financial support from the Slovenian Research Agency (AM, BL and MČ research core funding No. P1-0192, DL research core funding No. P2-0089 and AM and DL the project No. J7-8267). Also, the financial support from OeAD and the Slovenian Research Agency for bilateral project (WTZ grant number SI 13/2016 and ARRS grant number BI-AT/18-19-009) is acknowledged. We thank the CENN Nanocenter for the use of the LakeShore 7400 Series VSM vibrating-sample magnetometer and TEM Jeol 2100.
This work is based on experiments performed at the Swiss spallation neutron source SINQ, Paul Scherrer Institute, Villigen, Switzerland.

## Notes and references

1 S. Odenbach, *Colloidal Magnetic Fluids: Basics, Development and Application of Ferrofluids*, Springer, Berlin Heidelberg, 2009.
2 M. Shuai, A. Klittnick, Y. Shen, G. P. Smith, M. R. Tuchband, C. Zhu, R. G. Petschek, A. Mertelj, D. Lisjak, M. Čopič, J. E. Maclennan, M. A. Glaser and N. A. Clark, *Nat Commun*, 2016, **7**, 10394.
3 B. Huke and M. Lücke, *Rep. Prog. Phys.*, 2004, **67**, 1731–1768.
4 S. H. L. Klapp, *J. Phys.: Condens. Matter*, 2005, **17**, R525–R550.
5 A. Leushin and M. Eremin, *Zhurnal Eksperimentalnoi Teor. Fiz.*, 1975, **69**, 2190–2198.
6 K. Yamamoto, C. R. Hogg, S. Yamamuro, T. Hirayama and S. A. Majetich, *Appl. Phys. Lett.*, 2011, **98**, 072509.
7 K. I. Morozov, *The Journal of Chemical Physics*, 2003, **119**, 13024–13032.
8 S. H. L. Klapp and G. N. Patey, *The Journal of Chemical Physics*, 2001, **115**, 4718–4731.
9 F. M. van der Kooij, K. Kassapidou and H. N. W. Lekkerkerker, *Nature*, 2000, **406**, 868–871.
10 J.-C. P. Gabriel and P. Davidson, *Adv. Mater.*, 2000, **12**, 9–20.
11 D. van der Beek and H. N. W. Lekkerkerker, *EPL*, 2003, **61**, 702.
12 E. Paineau, A. M. Philippe, K. Antonova, I. Bihannic, P. Davidson, I. Dozov, J. C. P. Gabriel, M. Impéror-Clerc, P. Levitz, F. Meneau and L. J. Michot, *Liquid Crystals Reviews*, 2013, **1**, 110–126.
13 P. Davidson, C. Penisson, D. Constantin and J.-C. P. Gabriel, *PNAS*, 2018, 201802692.
14 F. Lin, X. Tong, Y. Wang, J. Bao and Z. M. Wang, *Nanoscale Research Letters*, 2015, **10**, 435.
15 M. Kotlarchyk and S. Chen, *The Journal of Chemical Physics*, 1983, **79**, 2461–2469.
16 G. L. Squires, *Introduction to the Theory of Thermal Neutron Scattering*, Dover Publications, Mineola, N.Y, 1997.
17 S. Mühlbauer, D. Honecker, É. A. Périgo, F. Bergner, S. Disch, A. Heinemann, S. Erokhin, D. Berkov, C. Leighton, M. R. Eskildsen and A. Michels, *Rev. Mod. Phys.*, 2019, **91**, 015004.
18 Wagner, W. and Kohlbrecher, J., in *Modern Techniques for Characterizing Magnetic Materials*, ed. Y. Zhu, Springer US, Boston, MA, 2005, pp. 65–105.
19 P. G. de Gennes and J. Prost, *The Physics of Liquid Crystals*, Clarendon Press, Oxford, 2nd edn., 1995.
20 P. Davidson, D. Petermann and A. M. Levelut, *J. Phys. II France*, 1995, **5**, 113–131.
21 R. Agra, E. Trizac and L. Bocquet, *Eur. Phys. J. E*, 2004, **15**, 345–357.
22 S. Ovtar, D. Lisjak and M. Drofenik, *Journal of Colloid and Interface Science*, 2009, **337**, 456–463.
23 D. Lisjak, S. Ovtar and M. Drofenik, *Journal of Materials Science*, 2011, **46**, 2851–2859.
24 M. Vis, H. H. Wensink, H. N. W. Lekkerkerker and D. Kleshchanok, *Molecular Physics*, 2015, **113**, 1053–1060.
25 H. H. Wensink and H. N. W. Lekkerkerker, *Molecular Physics*, 2009, **107**, 2111–2118.
26 A. Wiedenmann, A. Hoell, M. Kammel and P. Boesecke, *Phys. Rev. E*, 2003, **68**, 031203.
27 A. Wiedenmann, M. Kammel, A. Heinemann and U. Keiderling, *J. Phys.: Condens. Matter*, 2006, **18**, S2713.
28 O. Kuznetsova and A. Ivanov, *Journal of Magnetism and Magnetic Materials*, 2005, **289**, 226–229.
29 D. Lisjak and M. Drofenik, *Crystal Growth & Design*, 2012, **12**, 5174–5179.
30 D. Makovec, B. Belec, T. Goršak, D. Lisjak, M. Komelj, G. Dražić and S. Gyergyek, *Nanoscale*, 2018, **10**, 14480–14491.
31 D. Makovec, M. Komelj, G. Dražić, B. Belec, T. Goršak, S. Gyergyek and D. Lisjak, *Acta Materialia*, 2019, **172**, 84–91.
32 W. B. Russel, D. A. Saville and W. R. Schowalter, *Colloidal Dispersions*, Cambridge University Press, New Ed., 1992.